# Lustre, Hadoop, Accumulo

Jeremy Kepner[1,2,3], William Arcand[1], David Bestor[1], Bill Bergeron[1], Chansup Byun[1], Lauren Edwards[1], Vijay Gadepally[1,2], Matthew Hubbell[1], Peter Michaleas[1], Julie Mullen[1], Andrew Prout[1], Antonio Rosa[1], Charles Yee[1], Albert Reuther[1]

[1]MIT Lincoln Laboratory, [2]MIT Computer Science & AI Laboratory, [3]MIT Mathematics Department

*Abstract*—Data processing systems impose multiple views on data as it is processed by the system. These views include spreadsheets, databases, matrices, and graphs. There are a wide variety of technologies that can be used to store and process data through these different steps. The Lustre parallel file system, the Hadoop distributed file system, and the Accumulo database are all designed to address the largest and the most challenging data storage problems. There have been many ad-hoc comparisons of these technologies. This paper describes the foundational principles of each technology, provides simple models for assessing their capabilities, and compares the various technologies on a hypothetical common cluster. These comparisons indicate that Lustre provides 2x more storage capacity, is less likely to loose data during 3 simultaneous drive failures, and provides higher bandwidth on general purpose workloads. Hadoop can provide 4x greater read bandwidth on special purpose workloads. Accumulo provides $10^5$ lower latency on random lookups than either Lustre or Hadoop but Accumulo's bulk bandwidth is 10x less. Significant recent work has been done to enable mix-and-match solutions that allow Lustre, Hadoop, and Accumulo to be combined in different ways.

## I. INTRODUCTION

As data moves through a processing system the data are viewed from different perspectives by different parts of the system (see Figure 1). Data often are first parsed from a raw form (e.g., .json, .xml) into a tabular spreadsheet form (e.g., .csv or .tsv files), then ingested into database tables, analyzed with matrix mathematics, and presented as graphs of relationships. There are a wide variety of technologies that can be used to store and process data through these different steps. Three open source technologies of particular interest are the Lustre parallel file system (lustre.org) [Braam 2004], the Hadoop distributed file system and its map-reduce computation environment (hadoop.apache.org)[Bialecki et al 2005], and the Accumulo key-value database (accumulo.apache.org) [Wall, Cordova & Rinaldi 2013]. Each of these technologies is designed to solve particular challenges. This paper will review the design of each of these technologies, describe the kinds of problems they are designed to solve, and present some basic performance estimates for each. Finally, there are a number of ways these technologies can be combined, and they are also discussed.

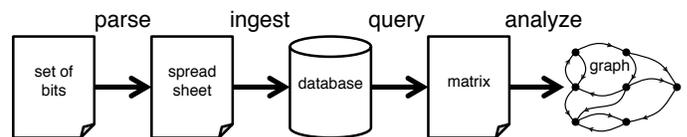

Figure 1. The standard steps in a data processing system often require different perspectives on the data.

## II. LUSTRE

Lustre is designed to meet the highest bandwidth file requirements on the largest systems in the world. The open source Lustre parallel file system presents itself as a standard POSIX, general purpose file system and is mounted by client computers running the Lustre client software. A file stored in Lustre is broken into two components: metadata and object data (see Figure 2). Metadata consists of the fields associated with each file such as filename, file permissions, and timestamps. Object data consists of the binary data stored in the file. File metadata is stored in the Lustre metadata server (MDS). Object data is stored in object storage servers (OSSes). When a client requests data from a file, the MDS returns pointers to the appropriate objects in the OSSes. This is transparent to the user and handled by the Lustre client. To an application, Lustre operations appear as standard file system operations and require no modification of application code.

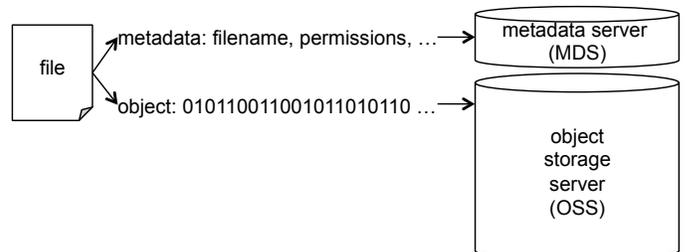

Figure 2. The Lustre file system splits a file into metdata and object data. The metadata is stored on the metadata server (MDS). The object data is store on the object storage server (OSS).

A typical Lustre installation might have two MDS servers (one active and one passive) that allows for hot failover. On each MDS, there might be several disk drives to hold the metadata. These drives are often formatted in a RAID10 configuration to allow any drive to fail without degrading performance.

A typical Luster installation might have many OSSes. In

This material is based upon work supported by the National Science Foundation under Grant No. DMS-1312831. Any opinions, findings, and conclusions or recommendations expressed in this material are those of the author(s) and do not necessarily reflect the views of the National Science Foundation.

turn, each OSS can have a large number of drives that are often formatted in a RAID6 configuration to allow for the failure of any two drives in an OSS. The many drives in an OSS allows data to be read in parallel at high bandwidth. File objects are striped across multiple OSSes to further increase parallel performance.

The above redundancy is designed to give Lustre high availability and no single point of failure. Data loss can only occur if three drives fail in the same OSS prior to any one of the failures being corrected. This probability is given by:

$P_3 = (n_d P_1)((n_d/n_s -1) P_1) ((n_d/n_s -2) P_1) \approx (n_d^3/n_s^2) P_1^3$,

where

$P_3$ = probability that 3 drives fail in the same OSS
$P_1$ = probability that a single drive fails
$n_d$ = number of drives in the entire system
$n_s$ = number of object storage servers

For a typical system with $n_s = 10$, then $P_3 = (n_d P_1)^3 / 100$.

The typical storage penalty for this redundancy is 35%. Thus, a system with 6 petabytes of raw storage might provide 4 petabytes of data capacity to its users.

Lustre is designed to deliver high read and write performance for many simultaneous large files. This is achieved by the clients having a direct connection to the OSSes via a high speed network. This connection is brokered by the MDS. The peak bandwidth of Lustre is determined by the aggregate network bandwidth to the client systems, the bisection bandwidth of the network switch, the aggregate network connection to the OSSes, and the aggregate bandwidth of the all the disks. More precisely, total Lustre bandwidth is given by

$B^{-1} = (n_c B_c)^{-1} + (B_n)^{-1} + (n_s B_s)^{-1} + (n_d B_d)^{-1}$

where

$B$ = total Luster bandwidth
$n_c$ = number of clients
$B_c$ = bandwidth of each client
$B_n$ = bisection bandwidth of the network switch
$n_s$ = number of object storage servers
$B_s$ = bandwidth of each object storage server
$n_d$ = number of drives in the system
$B_d$ = bandwidth of each drive

Consider a typical Lustre system where

$n_c$ = 100
$B_c$ = 1 GB/sec
$B_n$ = N/A (i.e., 1:1 non-blocking switch)
$n_s$ = 10
$B_s$ = 4 GB/s
$n_d$ = 1000
$B_d$ = 0.1 GB/sec
$\Rightarrow B$ = 22 GB/sec

Like most file systems Lustre is designed for sequential read access and not random lookups of data. To find any particular data value in Lustre requires on average scanning through half the file system. For the system described above with 4 petabytes of user storage, this would require ~1 day.

Finally, the Lustre security model is standard unix permissions.

### III. HADOOP

Hadoop is a fault-tolerant, distributed file system and distributed computation system. The Hadoop distributed file system (HDFS) is modeled after the Google File System (GFS)[Ghemawat et al 2003] and is a scalable distributed file system for large distributed data-intensive applications. GFS provides fault tolerance while running on inexpensive commodity hardware, and it delivers high aggregate performance to a large number of clients. The Hadoop distributed computation system uses the map-reduce parallel programming model for distributing computation onto the data nodes.

The foundational assumptions of HDFS are that its hardware and applications have the following properties [HDFS 2015]: high rates of hardware failures, special purpose applications, large data sets, write-once-read-many data, and read dominated applications. HDFS is designed for an important, but highly specialized class of applications for a specific class of hardware. In HDFS, applications primarily employ a co-design model whereby the HDFS file system is accessed via specific calls associated with the Hadoop API.

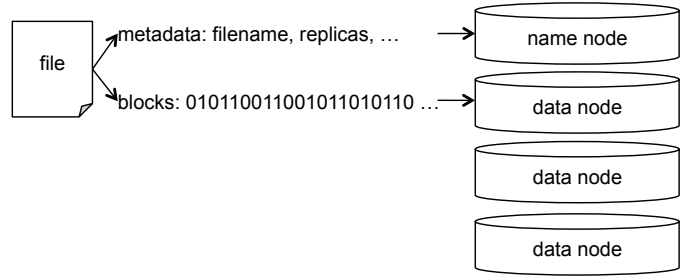

Figure 3. HDFS splits a file into metdata and replicated data blocks. The metadata is stored on the name node. The data blocks are stored on the data nodes.

A file stored in HDFS is broken into two components: metadata and data blocks (see Figure 3). Metadata consists of various fields such as the filename, creation date, and the number of replicas. Data blocks consist of the binary data stored in the file. File metadata is stored in the HDFS name node. Block data is stored on data nodes. HDFS is designed to store very large files that will be broken up into multiple data blocks. In addition, HDFS is designed to support fault-tolerance in massive distributed data centers. Each block has a specified number of replicas that are distributed across different data nodes.

"The placement of replicas is critical to HDFS reliability and performance. Optimizing replica placement distinguishes HDFS from most other distributed file systems. This is a feature that needs lots of tuning and experience." [HDFS R121]

One common HDFS replication policy is to store three copies of each data block in a location aware manner so that one copy is in another rack in the data center and that a second copy is in another data center. With this policy, the data will be protected from both rack unavailability and data center unavailability.

The storage penalty for a triple replication policy is 66%. Thus, a system with 6 petabytes of raw storage might provide 2 petabytes of data capacity to its users.

Data loss can only occur if three drives fail prior to any one of the failures being corrected. This probability is given by:

$P_3 = (n_d P_1)((n_d -1) P_1) ((n_d -2) P_1) \approx (n_d P_1)^3$

Hadoop is Java software that is typically installed in a special Hadoop user account and runs various Hadoop deamon processes to provide services to its clients. Hadoop applications contain special API calls to access the HDFS services. A typical Hadoop application using the map-reduce programming model will distribute an application over the file system so that each application is exclusively reading blocks that are local to the node that it is running on. A well-written Hadoop application can achieve very high performance if the blocks of the files are well distributed across the data nodes. Hadoop applications use the same hardware for storage and computation. The bandwidth achieved out of HDFS is highly dependent upon the computation to communication ratio of the Hadoop application. The total Hadoop read bandwith for optimally placed data blocks is given by

$B_{write} = \min(n_c, n_d) B_d / R$

$B_{read} = \min(n_c, n_d) B_d / (1 + r)$

where

$B$ = total HDFS bandwidth
$n_c$ = number of clients
$R$ = replication
$r$ = local compute time to read time ratio
$n_d$ = number of disks
$B_d$ = bandwidth of each disk

Consider a typical HDFS system where

$n_c$ = 1000
$R$ = 3
$n_d$ = 1000
$B_d$ = 0.1 GB/sec

$\Rightarrow B_{write}$ = 33 GB/sec

The compute time to read time ratio of a Hadoop application has a wide range. HDFS is designed for read dominated applications. For example, counting the number of words in a file involves reading all the data in the file and counting the number of bytes corresponding to the word separator characters. In this case, the local compute time to read time ratio $r \approx 0 \Rightarrow B_{read}$ = 100 GB/sec. Likewise, an application that sorts and parses words in a complex manner may have local compute time to read time ratio $r \approx 1000 \Rightarrow B_{read}$ = 100 MB/sec.

Like most file systems, HDFS is designed for sequential read access and not random lookups of data. To find any particular data value in HDFS requires scanning through half the file system. For the system described above with 4 petabytes of user storage, this would require ~3 hours assuming there is no other compute load on the Hadoop processors.

Finally, the HDFS security model is migrating towards standard unix permissions.

## IV. ACCUMULO

Relational or SQL (Structured Query Language) databases [Codd 1970, Stonebraker et al 1976] have been the de facto interface to databases since the 1980s and are the bedrock of electronic transactions around the world. More recently, key-value stores (NoSQL databases) [Chang et al 2008] have been developed for representing large sparse tables to aid in the analysis of data for Internet search. As a result, the majority of the data on the Internet is now analyzed using key-value stores [DeCandia et al 2007, Lakshman & Malik 2010, George 2011]. In response to the same challenges, the relational database community has developed a new class of array store (NewSQL) databases [Stonebraker et al 2005, Kallman et al 2008, Stonebraker & Weisberg 2013] to provide the features of relational databases while also scaling to very large data sets.

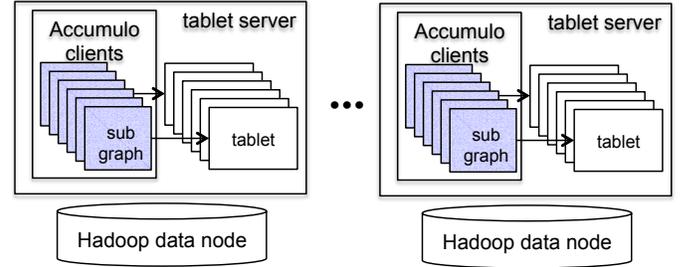

Figure 4. Accumulo is ideally suited for storing large graphs in tables. Tables are split into tablets that are hosted on tablet servers that reside on Hadoop data nodes. Accumulo client processes then can access different portions of the graph in parallel.

Accumulo is a unique NoSQL database that it is designed for the highest possible performance and scalability while also providing security labels for every entry in the database. Accumulo uses HDFS as its storage system. All Accumulo internal read and write processes are scheduled out of its own thread pool that is managed independently of the Hadoop map-reduce scheduler. Accumulo's main dependency on Hadoop is HDFS.

Accumulo is designed to run on large clusters of computing hardware where each node in the cluster has its own data storage. Accumulo uses the Hadoop Distributed File System (HDFS) to organize the storage on the nodes into a single, large, redundant file system (see Figure 4). A table in Accumulo is broken up into tablets where each tablet contains a continuous block of rows. The row values marking the boundaries between tablets are called *splits*. A table can be broken up into many tablets, and these tablets are then stored in HDFS across the cluster. Good performance is achieved when the data and the operations are spread evenly across the cluster. The selection of good splits is key to achieving this goal.

The various Accumulo processes are managed by Zookeeper (zookeeper.apache.org), which is a centralized service for maintaining configuration and naming information, along with providing distributed synchronization and group services. Accumulo's data redundancy relies on the underlying HDFS replication. Accumulo's availability relies on a heartbeat mechanism by which each Accumulo tablet server regularly reports in to Zookeeper. If a tablet server fails to report, Accumulo assumes the tablet server is unavailable and will not attempt to read from or write to the tablet server.

Accumulo is a key-value store where each entry consists of a seven-tuple. Most of the concepts of Accumulo can be understood by reducing this seven-tuple into a triple consisting of a row, column, and value. Each triple describes a point in a table. Only the non-empty entries are stored in each row, so the table can have an almost unlimited number of rows and columns and be extremely sparse, which makes Accumulo well suited for storing graphs.

The performance of Accumulo (and many other databases) is most commonly measured in terms of the rate at which entries can be inserted into a database table. A detailed mathematical model of Accumulo insert performance is beyond the scope of this paper. The peak performance of Accumulo has been measured at over 100,000,000 entries per second [Kepner et al 2014]. The peak insert rate for a single thread is typically ~100,000 entries per second. A typical single node server can reach ~500,000 entries per second using several insert threads [Sawyer 2013]. For the hypothetical Hadoop cluster described in the previous section, the peak performance would be ~100,000,000 entries per second. If a typical entry contains 30 bytes of data, this corresponds to a peak insert bandwidth of 3 Gigabytes/sec.

The Accumulo security model is a unique algebra that allows each data entry to be labeled so that a variety of tests can be done to determine if the data should be returned to the user.

V. HADOOP ON LUSTRE

As the Hadoop community has grown, it has become an increasingly popular API for a variety of applications. Applications written to the Hadoop map-reduce API now represent "legacy" applications in many organizations. Likewise, applications written to traditional file systems are also extremely popular and thus there has been strong interest in determining how these applications can coexist on shared computing hardware [Rutman 2011, Kulkarni 2013, DDN 2013, System Fabric Works 2014, Seagate 2014, Seagate 2015]. There are two primary approaches to running Hadoop applications on Lustre: porting map-reduce applications to Hadoop and swapping the underlying Hadoop file system for Lustre.

The first approach is to rewrite the Hadoop applications to use the map-reduce parallel programming model without using the Hadoop API [Byun et al 2012, Byun et al 2015]. This process mostly consists of removing the Hadoop API calls from the application and replacing them with regular file system calls. The application is then run using any of a number of standard schedulers. At the prices of porting the code, the net result of this process is usually applications that are smaller, easier to maintain, and run faster.

The second approach is modify the Lustre client with a plugin that replaces, or augments, the default Hadoop file system with the Lustre File System, which writes to a shared Lustre mount point that is accessible by all machines in the Hadoop cluster [Seagate 2015].

VI. ACCUMULO ON LUSTRE

Accumulo and Lustre are both designed to address the biggest and most challenging database and storage problems in their respective domains. Their common focus on performance and scalability via well-supported open source projects naturally brings these two communities together. Accmulo's relatively narrow dependence on HDFS means Accumulo is somewhat independent of its underlying storage environment, and applications don't need to be changed if underlying storage changes. Currently, there are two primary ways for Accumulo and Lustre to coexist: HDFS on Lustre and checkpoint/restart on Lustre.

As discussed in the previous section, running HDFS and Hadoop map-reduce on Lustre has been demonstrated by a number of organizations and is an available product. The Accumulo database has its own set of processes that run outside of Hadoop map-reduce. Accumulo's high availability is achieved by using heartbeat based fault-tolerance policies that can make Accumulo more susceptible to global system availability pauses then a standard Hadoop map-reduce job. Global system availability is a pressing concern in resource-constrained environments such a cloud VM environments. In such environments, as resource limits are approached, the system experiences pauses that can trigger Accumulo's fault tolerance mechanism to be invoked globally. Recent work by the Accumulo community [Fuchs 2015] has made significant progress for diagnosing when Accumulo is approaching resource limits (e.g., RAM limits) and to more gracefully handle its processes as resource limits are approached. The exact same issue can also be a concern for a Lustre system. If the Lustre MDS becomes heavily loaded, it may momentarily respond more slowly which can invoke the Accumulo fault-tolerance mechanism. The recent efforts to improve Accumulo behavior on VMs are also applicable to improving Accumulo running on Lustre.

A typical large Accumulo instance is physically coupled to its underlying compute system. Starting, stopping, checkpoint, cloning, and restarting an Accumulo instance on different hardware is difficult. Likewise keeping many production scale Accumulo instances available for database developers can also be challenging. Lustre can play an obvious role in this setting by being a repository by which entire Accumulo instances can be stored while they are not running. The MIT SuperCloud [Reuther et al 2013, Prout et al 2015] adopts this approach and it has proven to be very useful for developers and for managing production Accumulo instances.

VII. SUMMARY

The Lustre parallel file system, the Hadoop distributed file system, and the Accumulo database are all designed to address the largest and the most challenging data storage problems. There have been many ad-hoc comparisons of these technologies. This paper describes the foundational principles of each technology, provides simple models for assessing their capabilities, and compares the various technologies on a hypothetical common cluster. These comparisons are shown in Table 1 and indicate that the Lustre provides 2x more storage capacity, is less likely to loose data during 3 simultaneous drive failures, and provides higher bandwidth across general purpose workloads. Hadoop can provide 4x greater read bandwidth on special purpose workloads. Accumulo provides $10^5$ lower latency on random lookups than either Lustre or Hadoop but Accumulo's overall bandwidth is 10x less. Significant recent work has been done to enable mix-and-match solutions that allow Lustre, Hadoop, and Accumulo to be combined in different ways.

Table 1. Comparison of the peak capabilities of different technologies for the hypothetical system described in the previous sections.

| Technology | Total Storage | Loss Prob | File Write | File Read | Random Lookup |
|---|---|---|---|---|---|
| Lustre | 4 PB | $(n_d P_1)^3/100$ | 22 GB/s | 22 GB/s | 1 day |
| Hadoop | 2 PB | $(n_d P_1)^3$ | 33 GB/s | 100* GB/s | 3* hours |
| Accumulo | 2 PB | $(n_d P_1)^3$ | 3 GB/s | 3 GB/s | 50 msec |
| Hadoop/Lustre | 4 PB | $(n_d P_1)^3/100$ | 22 GB/s | 22 GB/s | 1 day |
| Accmulo/Lustre | 4 PB | $(n_d P_1)^3/100$ | 3 GB/s | 3 GB/s | 50 msec |

*Assumes perfect data placement and no other processes on clients.